\documentclass{PoS}
\usepackage[utf8]{inputenc}
\usepackage{amsmath,amsthm,amsfonts,amssymb,graphicx}

\title{Generalising Courant algebroids to M-theory}

\ShortTitle{Generalising Courant algebroids to M-theory}

\author{\speaker{Alex S.~Arvanitakis}\\
        \small \em The Blackett Laboratory,\\
\small \em Imperial College London, \\
\small \em Prince Consort Road, London SW7 2AZ, U.K.\\
        E-mail: \email{a.arvanitakis@imperial.ac.uk}}

\abstract{We introduce M-theoretic generalisations of the notion of (exact) Courant algebroid, and summarise their connections to generalised geometry, U-duality, and the physics of strings, membranes, and fivebranes.

This is a summary of the paper \cite{Arvanitakis:2018cyo}, presented at CORFU2018.}

\FullConference{Corfu Summer Institute 2018 "School and Workshops on Elementary Particle Physics and Gravity"\\
		(CORFU2018)\\
		31 August - 28 September, 2018\\
		Corfu, Greece}

\begin{document}

In this contribution we will explain the relationships encapsulated in the following table:

\begin{table}[h]\centering
\begin{tabular}{c l l l c}
Group & Gen.~Tangent bundle & SUGRA &Algebroid & Brane\\
\hline
$O(d,d\,;\mathbb R)$ & $T\oplus T^\star$ & Type II  & Exact Courant &  String\\
$SL(5\,;\mathbb R)$ &$T\oplus \Lambda^2 T^\star$ & 11d& ``M2'' & Membrane\\
$E_{6(6)}$ &$T\oplus \Lambda^2 T^\star\oplus \Lambda^5 T^\star$ & 11d& ``M5'' & Fivebrane \\
\end{tabular}
\end{table}
The first row summarises the well-known fact that the exact Courant algebroid provides the mathematical setting for the description of type II supergravity\footnote{the Neveu-Schwarz (NS) sector thereof, actually. We ignore the dilaton because it is a (generalised) scalar.} backgrounds in terms of Hitchin-Gualtieri generalised geometry \cite{Gualtieri:2003dx,Hitchin:2004ut}. The group $O(d,d\,;\mathbb R)$ acts on such backgrounds --- ``background'' here meaning a Lorentzian metric $g$ and Kalb-Ramond field $b\in \Gamma[\Lambda^2 T^\star ]$ on a $d$-dimensional manifold $M$  (where $T,T^\star$ are the tangent and cotangent bundles of $M$, respectively) --- and that action naturally arises from string theory: backgrounds related by the action of the discrete subgroup $O(d,d\,;\mathbb Z)$ define physically equivalent string theories. This equivalence is known as \emph{T-duality}. The $O(d,d\,;\mathbb R)$-action on $(g,b)$ is best described through the introduction of a \emph{generalised metric} $\mathcal H$:
\begin{align}
\mathcal H=\begin{pmatrix} g- b g^{-1}b & bg^{-1} \\ -g^{-1}b & g^{-1}\end{pmatrix}
\end{align}
which is a symmetric bilinear form on the \emph{generalised tangent bundle} $E$
\begin{align}
E\equiv T\oplus T^\star
\end{align}
on which the $O(d,d\,;\mathbb R)$-action is obvious. Infinitesimal diffeomorphisms and gauge transformations of $(g,b)$ induce the action of a \emph{generalised Lie derivative} or \emph{Dorfman bracket} $L_A$ (with $A$ a section of $T\oplus T^\star$) on $\mathcal H$, which can be calculated by the Leibniz rule from the definition
\begin{align}
L_A A'\equiv (\mathcal L_v v'\,,\; \mathcal L_v\lambda' - \iota_{v'}d\lambda)\in\Gamma[E]\,,\qquad A=(v,\lambda)\,,\;A'=(v',\lambda')\in\Gamma[E]
\end{align}
where $v,v'$ are vector fields and $\lambda,\lambda'$ are 1-forms on $M$ respectively, while $\mathcal L_v$ and $d$ are the Lie derivative and exterior derivative.

The vector bundle $E=T\oplus T^\star$ with the natural $O(d,d\,;\mathbb R)$ metric $\eta$, projection $E\to T$ and  Dorfman bracket $L_{\--}$ forms the canonical example of a Courant algebroid. A Courant algebroid \cite{liu1995manin} is defined by the data $(E,\eta, L_{\--},\rho)$ of a vector bundle $E$, nondegenerate bundle metric $\eta$, Dorfman bracket $L_{\--}\--: \Gamma[E]\times \Gamma[E]\to \Gamma[E]$ and bundle map $\rho:E\to T$ satisfying the axioms
\begin{align}
L_{A}(f A')&= f L_{A} A'+ (\rho(A)\cdot f) A'\,,\qquad f\in C^\infty(M)\,,\\
\rho(L_{A}A')&=\mathcal L_{\rho(A)}\rho(A')\\
\rho(A)\cdot \eta(B,C)&=\eta(L_{A}B,C)+\eta(B,L_A C)\,,\qquad A,B,C\in\Gamma[E]\\
L_{A}(L_B C)&=L_{L_A B}C + L_B(L_A C) \label{leibniz}\\
L_A A&=\frac{1}{2}\rho^\star_\eta\big (d (\eta(A,A))\big)\,, \qquad\eta(\rho^\star_\eta(\lambda),A)\equiv \langle \lambda, \rho(A)\rangle\,,\quad\lambda\in \Gamma[T^\star]\,.
\end{align}
We have here given the definition of \cite{Severa:2017oew} because the Dorfman bracket is of more direct physical relevance (the usual definition involves a ``Courant bracket'' which is the antisymmetric part of the Dorfman bracket). An important implication here is that $\rho\circ \rho^\star_\eta=0$ so the sequence
\begin{align}
0\to T^\star \xrightarrow{\rho^\star_\eta} E\xrightarrow{\rho} T\to 0
\end{align}
is a chain complex of vector bundles. If this is an exact sequence this is called an \emph{exact} Courant algebroid. In that case the fibre $E_x\cong T_x\oplus T^\star_x$ over every point $x\in M$, so $E$ locally looks like the motivating example of $T\oplus T^\star$. Hitchin-Gualtieri generalised geometry \cite{Gualtieri:2003dx,Hitchin:2004ut} largely involves the study of the vector bundle $E$ of an exact Courant algebroid  and geometric structures on it, as opposed to structures on the ordinary tangent bundle $T$.

Exact Courant algebroids are classified by the 3rd de Rham cohomology $H^3(M,\mathbb R)$ \cite{Severa:2017oew} i.e.~by a closed 3-form up to exact ones. This can in fact be identified with the closed 3-form field strength\footnote{We may call this the ``NS 3-form'' later.} $H\in\Gamma[\Lambda^3 T^\star]$ (up to gauge transformations) of the Kalb-Ramond field $b$ appearing in type II backgrounds. We will see shortly that this identification is realised by the canonical association of a certain topological field theory to the data of a Courant algebroid, which is equivalent to the following interaction between a string of worldsheet $\Sigma$ propagating on the spacetime $M$ and the 3-form $H$
\begin{align}
\int_{Y} x^\star(H)\,,\qquad x:Y\to M\,,\quad \partial Y=\Sigma
\end{align}
where $Y$ is any 3-manifold with boundary the worldsheet $\Sigma$, and $x$ is the map describing the immersion of the string worldsheet $\Sigma$ in spacetime $M$, extended to $Y$. This is known as a ``Wess-Zumino term''.

\bigskip

It is remarkable that this entire picture generalises to eleven-dimensional supergravity (11d SUGRA) --- i.e.~the low-energy limit of M-theory, which we may conflate here --- as suggested by the table: there exist algebroids generalising the exact Courant algebroid, which form the mathematical setting for the \emph{exceptional} generalised geometry \cite{Hull:2007zu,Pacheco:2008ps} relevant for M-theory backgrounds of the product form
\begin{align}
\mathbb R^{11-d}\times M \quad \text{or}\quad {\rm AdS}^{11-d}\times M \label{compactifications}
\end{align}
where $M$ is a $d$-dimensional manifold. The moduli of such backgrounds can be packaged into tensors over generalised tangent bundles $E$ as in the table; in particular the bosonic fields $(g,C)$ namely the metric $g$ on $M$ (now usually taken to be Riemannian) along with the SUGRA 3-form potential $C$ are packaged into a generalised metric which transforms nicely under (continuous versions of) the U-duality groups $E_{d(d)}$: the split real forms of the exceptional Lie groups $E_8,\,E_7,\,E_6$, continued for lower $d$ to ${\rm Spin}(5,5),\, SL(5),\, SL(3)\times SL(2),\,SL(2)\times \mathbb R^+$. Said algebroids are classified by certain classes which can be identified with gauge equivalence classes of $C$ and its dual, and have associated topological field theories that yield the Wess-Zumino terms for the interaction between the fundamental M-theory branes
 --- the membrane (M2) and fivebrane (M5) respectively --- and the $C$ field.\footnote{Actually, there is a similar picture for type IIB supergravity including Ramond-Ramond fields; there is a ``D3''-algebroid \cite{Arvanitakis:2018cyo}.}

 We will proceed to describe all three algebroids of the table at once using the language of NPQ manifolds (used also in e.g.~the original AKSZ paper \cite{Alexandrov:1995kv}), which is how the M2 algebroid (which appeared in 2010 in \cite{Ikeda:2010vz}) and M5 algebroid (in \cite{Arvanitakis:2018cyo}) were originally written down anyway. The relation of the M2 algebroid to $SL(5)$-exceptional generalised geometry was first mentioned in \cite{Kokenyesi:2018ynq}, while the general picture of algebroids and their associated branes sketched in the above paragraph in generality is one first main point of \cite{Arvanitakis:2018cyo} which this contribution is summarising. (The other points raised therein are the novel appearance of the worldvolume gauge field for the M5 which has no analogue in the string or M2 cases, and the mathematical interpretation of the ``tensor hierarchy'' in exceptional generalised geometry as an $L_\infty$-algebra obtained from the NPQ manifold structure.)

 \section{NPQ manifolds}
 An NPQ manifold is a supermanifold $\mathcal M$ which is in particular $\mathbb N$-graded (``N-structure'') with a homological vector field $Q$ (``Q-structure'') compatible with a graded symplectic form $\omega$ (``P-structure'') of some definite degree.

 The N-structure is the statement that the structure sheaf of $\mathcal M$ is a sheaf of $\mathbb Z$-graded rings with no generators of negative degree which are simultaneously $\mathbb Z_2$-graded-commutative such that parity is determined by the $\mathbb Z$ grading modulo 2. In particular the ring of functions $C(\mathcal M)$ of $\mathcal M$ is a $\mathbb Z$-graded ring, so it decomposes into subspaces $C_n(\mathcal M)$ of definite \emph{degree} or \emph{grading} $n\in\mathbb Z$, with $n\geq 0$ and furthermore
 \begin{align}
 f g =(-1)^{mn}g f \in C_{m+n}(\mathcal M)\,, \qquad \forall f\in C_{m}(\mathcal M)\,,\; g\in C_n(\mathcal M)\,.
 \end{align}
 In the mathematics literature these appeared in \cite{kontsevich2003deformation,Severa:2017oew}. Vector fields $V$ can then be defined as $\mathbb Z$-graded-derivations over $C(M)$:
 \begin{align}
 V(f)\in C_{\deg f+\deg V}(\mathcal M)\,,\quad V(fg)= V(f) g +(-1)^{(\deg V)(\deg f)}f V(g)
 \end{align}
 for $f,g \in C(\mathcal M)$ of definite degree; in particular the homological vector field $Q$ is a derivation of degree $1$ that squares to zero
 \begin{align}
 Q^2=0\,.
 \end{align}

A motivating example of an NQ-manifold is the shifted tangent bundle $T[1]M$ of an (ordinary \emph{or} super) manifold $M$ --- whose ring of functions $C(T[1]M)$ is isomorphic to the space of polyforms over $M$ --- and where $Q$ is identified with the de Rham exterior derivative. This identification is easy to see in local coordinates. Say $\{x^\mu\}$ are coordinates on a chart of $M$, then $\{x^\mu,dx^\nu\}$ are local coordinates of $T[1]M$ where $\deg dx^\mu=\deg x^\mu+1$, and $d(x^\mu)=dx^\mu$. A differential form of rank $p$ is then a degree-$p$ polynomial in $dx$.

This trick serves \emph{mutatis mutandis} to define differential forms over any NQ manifold $\mathcal M$. $T[1]\mathcal M$ can be defined as the space of maps ${\rm Maps}(\mathbb R[-1],\mathcal M)$ where $\mathbb R[-1]$ is the real linear $\mathbb Z$-graded supermanifold with a single odd coordinate $\theta$ of degree $-1$. $T[1]\mathcal M$ then carries two  gradings --- the original grading descended from that of $\mathcal M$ and the form grading --- which are compatible in the sense that the sum of the gradings define an N-structure on $T[1]\mathcal M$\footnote{This is easiest to understand if one introduces degree-counting vector fields (also known as ``Euler vector fields''). In local homogeneous coordinates $z^a$ on any N-manifold $\mathcal M$ the Euler vector field is $\epsilon=\deg(z^a) z^a\frac{\partial}{\partial z^a}$ and $f\in C_{n}(\mathcal M)\iff \epsilon\cdot f=n f$. $\epsilon$ also acts on $C(T[1]\mathcal M)$ by Lie derivative (and thus lifts to a vector field on $T[1]\mathcal M$) and thereby determines $\mathcal M$-degree of forms. However $T[1]\mathcal M$ also has the form degree counted by $\deg (dz^a)dz^a\frac{\partial}{\partial dz^a}$. The sum of the two gives the N-structure on $T[1]\mathcal M$.}. A graded symplectic form $\omega$ of degree $p$ on $\mathcal M$ is then an element $\omega\in C(T[1]\mathcal M)$ which has form degree $2$ and $\mathcal M$-degree $p$ (we will abbreviate this to simply ``degree $p$'') and is non-degenerate as a bilinear form.

An NPQ manifold $(\mathcal M,\omega,Q)$ of degree $p\geq 0$ is then an NQ manifold $(\mathcal M,Q)$ with degree $p$ symplectic form $\omega$ such that
\begin{align}
\mathcal L_Q\omega=0\,,
\end{align}
where $\mathcal L_Q\omega$ is the super Lie derivative of $\omega$ along $Q$. For $p>0$ which will be for us the case this implies the existence of a (globally defined!) hamiltonian function $\Theta\in C_{p+1}(\mathcal M)$:
\begin{align}
Q(f)=(\Theta,f)\,,\qquad \forall f \in C(\mathcal M)
\end{align}
where $(\--\,,\;\--)$ is the graded Poisson bracket associated to $\omega$, which carries $\mathcal M$-degree $-p$. Furthermore, we have
\begin{align}
\boxed{Q^2=0\iff (\Theta,\Theta)=0}
\end{align}
We refer to \cite{Arvanitakis:2018cyo} for details and proofs. Here we will only need to use the fact that the Poisson bracket is a graded left derivation on its right argument (like vector fields on supermanifolds are in our convention above) and a right derivation on its left argument.

\section{Algebroids from NPQ-manifolds}
Consider an NPQ manifold $\mathcal M$ of degree $p>0$. The spaces $C_n(\mathcal M)$ of functions of degree $n$ are clearly modules over $C_0(\mathcal M)\cong C(M)$ where we identified $C_0(\mathcal M)$ with the ring of functions $C(M)$ of the associated \emph{ordinary} manifold $M$ i.e.~the body of $\mathcal M$. In a local coordinate chart such that the symplectic form is constant (existence of such ``Darboux'' charts is easy and a proof is in e.g.~\cite{kajiura}), since degrees are non-negative and $\omega$ is non-degenerate we find that the maximum degree of a local coordinate is $p$. Therefore it is easy to see that each $C_n(\mathcal M)$ arises from a locally free sheaf of $C(M)$-modules of finite rank. By a smooth version of the Serre-Swan theorem (11.33 in \cite{nestruev2006smooth}) we find that each $C_n(\mathcal M)$ is the space of sections of a vector bundle of finite rank with base $M$.

(Some of) these vector bundles are where the algebroid structures will be living on. For the operations on them we use the derived bracket construction of Kosmann-Schwarzbach \cite{kosmann1996poisson} and Voronov \cite{voronov2005higher}. For $f\in C_m(\mathcal M)$ and $g \in C_n{(\mathcal M)}$ we can define a map $C_m(\mathcal M)\times C_n(\mathcal M)\to C_{m+n+1-p}(\mathcal M)$ by the formula (``derived bracket'')
\begin{align}
-\big((\Theta,f),g\big)=-\big( Q(f),g\big)
\end{align}
where the sign is just a convention. The virtue of this definition is that inequivalent algebroid structures are classified by the choice of $\Theta \in C_{p+1}(\mathcal M)$ up to symplectomorphisms of $(\mathcal M,\omega)$ as a $\mathbb Z$-graded symplectic supermanifold. It is also clear that algebraic properties of the operations follow immediately from $Q^2=0$ and the graded Jacobi/symmetry identities obeyed by the Poisson bracket $(\--,\,\--)$.

To illustrate the last point let us extract a (not necessarily (anti)-symmetric) bracket $L_{\--}\--:\Gamma[E]\times \Gamma[E]\to \Gamma[E]$ from the above derived bracket construction where $\Gamma[E]$ is identified with some space $C_q(\mathcal M)$ for some $q$. This is to be some kind of generalised Dorfman bracket (to draw an analogy with the Courant algebroid). Degree-counting implies in fact $q=p-1$. If we let $A,B,C\in C_{p-1}(\mathcal M)$ and define
\begin{align}
L_A B=- \big((\Theta,A),B\big) \label{derivedDorfman}
\end{align}
an easy calculation implies the identity \eqref{leibniz}! Said identity taken alone defines a ``Leibniz'' or ``Loday'' algebroid. Importantly, all algebroids of the table obey this identity.

Before we specialise to the algebroids of physical interest we will make this discussion explicit for the case of Vinogradov algebroids \cite{vinogradov1990union,grutzmann2015general} which are anyway an important ingredient in our constructions: take $\mathcal M=T^\star[p]T[1] M$, a shifted cotangent bundle of the shifted tangent bundle of an ordinary manifold $M$, and introduce a Darboux chart so that (for even $p$ henceforth for simplicity although everything works for odd $p$ too with different signs)
\begin{align}
\omega= dp_\mu dx^\mu- d\chi_\mu d\psi^\mu\implies (x^\mu,p_\nu)=-(p_\nu,x^\mu)\delta^\mu_\nu\,,\quad (\psi^\mu,\chi_\nu)=(\chi_\nu,\psi^\mu)=\delta^\mu_\nu
\end{align}
where $\deg(x^\mu,\psi^\mu,\chi_\mu,p_\mu)=(0,1,p-1,p)$ respectively. ($x^\mu$ form a local coordinate chart on $M$, while $(x^\mu,\psi^\mu)$ are a chart on $T[1]M$; in particular $\psi^\mu$ can be identified with $dx^\mu$ and the de Rham differential with $\psi^\mu\partial/\partial x^\mu$.) Clearly $\omega$ is a graded symplectic form of degree $p$. We have an NPQ manifold structure if we select a hamiltonian function $\Theta\in C_{p+1}(\mathcal M)$ such that $(\Theta,\Theta)=0$. {\bf We will assume that $\Theta$ has a term involving both $\psi$ and $p$ of the form $M^\mu_\nu p_\mu \psi^\nu$ with $M$ invertible, possibly $x$-dependent.}\footnote{This can be phrased invariantly in terms of surjectivity of the anchor map (an analogue of $\rho$ in this general case).} Given this constraint and the assumption it can be shown that any $\chi_\mu$ dependence in $\Theta$ can be removed by a symplectomorphism. Without loss of generality we can then write
\begin{align}
\Theta=  \psi^\mu p_\mu + \beta_{\mu_1\mu_2\cdots \mu_{p+1}}(x) \psi^{\mu_1}\cdots \psi^{\mu_{p+1}}\,.
\end{align}
In $\beta_{\mu_1\mu_2\cdots \mu_{p+1}}(x)$ we recognise the components of a $(p+1)$-form on $M$. We find $(\Theta,\Theta)=0$ if and only if $d\beta=0$. Since $Q=(\Theta,\--)$ acts as $\psi^\mu\partial/\partial x^\mu$ on anything that is $p_\mu$-independent, and $\exp( R,\--)$ is a symplectomorphism for $R\in C_p(\mathcal M)$ we find that inequivalent Vinogradov algebroids as above are classified by the de Rham cohomology class $H^{p+1}(M,\mathbb R)$. This procedure recovers the Vinogradov bracket on $C_{p-1}(\mathcal M)$ (in general ``twisted'' in the presence of a nontrivial $\beta$) which can be locally identified with $TM\oplus \Lambda^{p-1}T^\star M$: in the above chart $A\in C_{p-1}(\mathcal M)$ takes the form
\begin{align}
A= v^\mu(x) \chi_\mu + ((p-1)!)^{-1} \lambda_{\mu_1\mu_2\cdots \mu_{p-1}}(x) \psi^{\mu_1}\cdots \psi^{\mu_{p-1}}\,. \label{vinogradovgentanbundle}
\end{align}
For $\beta=0$, from \eqref{derivedDorfman} we indeed find the original form of the Vinogradov bracket on $TM\oplus \Lambda^{p-1}T^\star M$ \cite{vinogradov1990union}
\begin{align}
L_A A'=\mathcal L_{v}v' +\mathcal L_v \lambda' - \iota_{v'}d\lambda\,. \label{vinogradovbracket}
\end{align}

\section{Generalisations of Courant algebroids and string/M-theory
}
Returning to the table we notice the first two rows involve generalised tangent bundles of the Vinogradov form $TM\oplus \Lambda^{p-1}T^\star M$ for $p=2$ and $p=3$! It has in fact been known since the work \cite{Roytenberg:2002nu} that any Courant algebroid defines a $p=2$ NPQ manifold, and vice-versa. The $p=2$ NPQ manifold version of the Vinogradov algebroid we just discussed is simply the special case of this correspondence for \emph{exact} Courant algebroids. The bracket \eqref{vinogradovbracket} is indeed the generalised Lie derivative appearing in the physical literature on (exceptional) generalised geometry \cite{Hull:2007zu,Pacheco:2008ps} and encodes the action of infinitesimal diffeomorphisms and $b$ or $C$-field gauge transformations respectively. (Notice that the gauge parameter is respectively a $1$- or $2$-form and is therefore identified with a section of $\Lambda^{p-1}T^\star M$.)

We have already seen how the generalised Lie derivative of the exact Courant algebroid is obtained from the derived bracket expression \eqref{derivedDorfman}. We furthermore obtain however the map $\rho:E\to T$ appearing in the first definition of Courant algebroids by the same expression: for $\mathcal M=T^\star[2]T[1]M$ we have $A \in C_1(\mathcal M)$ taking the form \eqref{vinogradovgentanbundle} (so is locally a section of $T\oplus T^\star$). Then if $f\in C_0(\mathcal M)=C(M)$ we calculate
\begin{align}
\rho(A)\cdot f\equiv-\big((\Theta,A),f\big)=v^\mu\frac{\partial}{\partial x^\mu}f\,.
\end{align}
Finally, the $O(d,d)$ metric $\eta$ on the fibres of $E$ is obtained from the Poisson bracket restricted to functions of degree 1:
\begin{align}
(A,A')=(v^\mu(x)\chi_\mu + \lambda_\mu(x) \psi^\mu,\,v'^\nu(x)\chi_\nu + \lambda'_\nu(x) \psi^\nu)=v^\mu \lambda'_\mu + \lambda_\mu v'^\mu\,.
\end{align}

Following the pattern of the Courant algebroid and observing that the Poisson bracket is $C(M)$-linear in either argument when restricted to $C_n(\mathcal M)\,, n<p$ for a degree $p$ NPQ manifold $\mathcal M$ suggests that while all spaces $C_n(\mathcal M)$ correspond to sections of vector bundles, the ones which should feature in a definition of a kind of generalised Courant algebroid are the spaces of degree less than $p$. Therefore we can define the ``M2'' algebroid in terms of the $p=3$ Vinogradov algebroid as described in the previous section. For $\beta=0$ (i.e.~$\Theta=\psi^\mu p_\mu$) the relevant spaces $C_n(\mathcal M)$ are
\begin{align}
C_0(\mathcal M)=C(M)\,,\quad C_1(\mathcal M)\cong \Gamma[T^\star M]\,,\quad C_2(\mathcal M)\cong \Gamma[TM\oplus \Lambda^2T^\star M]
\end{align}
and the Poisson bracket and derived bracket \eqref{derivedDorfman} operations serve to define bilinear products and Dorfman-like brackets respectively involving these modules. It can be checked by comparison with the exceptional field theory literature\footnote{Exceptional generalised geometry as double field theory is to Hitchin-Gualtieri generalised geometry. We refer to the exceptional field theory literature here as opposed to the exceptional generalised geometry literature even though the latter would be more appropriate because the latter has not really been developed for smaller duality groups.}  that these operations correspond to the operations appearing in the ``tensor hierarchy'' therein --- roughly, the sequence of ``gauge for gauge'' transformations in addition to those parameterised by the generalised tangent bundle\footnote{There are discrepancies here related to the fact that our NPQ constructions are not fully U-duality-group covariant. This is to be expected, and is explained in \cite{Arvanitakis:2018cyo}.}. This is the case for $\dim M\leq 4$, for which $C_{2}(\mathcal M)$ has an action of the (continuous version of the) U-duality groups in the physically correct representations. For $\dim M=4$ this is the group $SL(5;\mathbb R)$ in the rank 2 antisymmetric tensor representation, whose dimension --- 10 --- indeed matches the rank of the vector bundle $TM\oplus \Lambda^2T^\star M$ for $\dim M=4$. Before moving on to the more interesting case of the M5 we recall that inequivalent algebroid structures are classified by a class $H^{4}(M,\mathbb R)$. This is highly suggestive of the 4-form field strength of the 11d SUGRA $C$-field 3-form gauge potential.
\bigskip

The novel case is that of the third row of the table: the M5 algebroid. This features prominently in $E_6$ exceptional generalised geometry of which a brief summary can be found in an appendix of e.g.~\cite{Ashmore:2015joa}. The generalised tangent bundle $E$ is locally isomorphic to
\begin{align}
TM\oplus \Lambda^2 T^\star M\oplus \Lambda^5 T^\star M \label{e6gentangent}
\end{align}
whose physical interpretation is that its sections are infinitesimal gauge parameters for the metric $g$ on $M$, the $C$-field, and its \emph{dual}. (In the context of an 11-dimensional spacetime of the form \eqref{compactifications} of which $M$ is a factor, if $G$ is the 11-dimensional metric (which induces $g$ on $M$) we can use the Hodge star associated to $G$ to form $\star dC$, which will be $d$-closed in the absence of sources. We can then use the Poincar\'e lemma to introduce the dual potential $\tilde C$ such that $d\tilde C=\star dC$. Pulling $\tilde C$ back to $M$ then leads to an independent potential (which we will also write $\tilde C$) which is a 6-form, and therefore has an associated 5-form gauge parameter, which accounts for the last summand in \eqref{e6gentangent}. (We have omitted a slight subtlety involving the Chern-Simons term of 11d SUGRA which shifts the definition of $\tilde C$.)) The fibres of this bundle form the 27-dimensional representation of the split real form of $E_6$ \cite{Hull:2007zu,Pacheco:2008ps} if $\dim M=6$, which is the case relevant for 6-dimensional compactifications of M-theory.

There is an exceptional Dorfman bracket or generalised Lie derivative $\Gamma[E]\times \Gamma[E]\to \Gamma[E]$ which obeys \eqref{leibniz}, rendering $E$ a Leibniz algebroid. We therefore see there is no Vinogradov algebroid that can possibly house this. However, we can generalise slightly as follows: take
\begin{align}
\mathcal M=T^\star[6]T[1]M\times \mathbb R[3]\,.
\end{align}
This is an NPQ manifold of degree 6 with hamiltonian $\Theta$ to be determined and symplectic structure
\begin{align}
\omega= dp_\mu dx^\mu- d\chi_\mu d\psi^\mu - \frac{1}{2}d\zeta d\zeta\\\implies (x^\mu,p_\nu)=-(p_\nu,x^\mu)\delta^\mu_\nu\,,\quad (\psi^\mu,\chi_\nu)=(\chi_\nu,\psi^\mu)=\delta^\mu_\nu\,,\quad (\zeta,\zeta)=1
\end{align}
where we have expressed $\omega$ in Darboux coordinates $(x^\mu,\psi^\mu,\zeta,\chi_\mu,p_\mu)$ whose degrees are respectively $(0,1,3,5,6)$. In particular $\zeta$ is the single odd coordinate parameterising $\mathbb R[3]$. We have used here the crucial fact that the space $\mathbb R[2n+1]\,, n\in\mathbb Z$ is in fact a $\mathbb Z$-graded symplectic vector space with the obvious symplectic structure of degree $4n+2$.

Clearly the space $C_{p-1=5}(\mathcal M)$ can accommodate sections of $M\oplus \Lambda^2 T^\star M\oplus \Lambda^5 T^\star M$: for $A\in C_5(\mathcal M)$ we find
\begin{align}
A= v^\mu(x) \chi_\mu + 2^{-1} \lambda_{\mu_1\mu_2}(x) \zeta \psi^{\mu_1}\psi^{\mu_2}+ (5!)^{-1} \sigma_{\mu_1\mu_2\cdots \mu_{5}}(x) \psi^{\mu_1}\cdots \psi^{\mu_{5}}
\end{align}
owing to the exceptional equality $2+3=5$. For $\Theta=\psi^\mu p_\mu$ (i.e.~in the absence of twists) the derived bracket expression \eqref{derivedDorfman} agrees with the following known formula for the exceptional Dorfman bracket
\begin{align}
L_A A'=\mathcal L_v v' + (\mathcal L_v \lambda'-\iota_{v'}d\lambda)+(\mathcal L_v \sigma'-\iota_{v'} d\sigma -\lambda' d\lambda)\,.
\end{align}
The novelty here is the very last term which is quadratic in sections of $\Lambda^2 T^\star$, for which the graded Poisson bracket $(\zeta,\zeta)=1$ is crucial. One can proceed to sketch a definition of an M5 algebroid involving the vector bundles associated to the spaces
\begin{align}
C_0(\mathcal M)=C(M)\,, C_1(\mathcal M)\,, C_2(\mathcal M)\,\dots C_5(\mathcal M)\,,
\end{align}
using the Poisson bracket and derived bracket \eqref{derivedDorfman} as before. In \cite{Arvanitakis:2018cyo} we detail how the spaces $C_{n<5}(\mathcal M)$ are naturally associated to the ``tensor hierarchy'' of exceptional generalised geometry/exceptional field theory, derived previously via ad hoc methods.

How many inequivalent M5 algebroids are there? Restricting to $\Theta=\psi^\mu p_\mu +\dots$ it is again trivial to show there are no terms depending on $\chi_\mu$ consistent with $(\Theta,\Theta)=0$ up to symplectomorphisms. Since $\Theta\in C_7(\mathcal M)$ in this case we arrive at
\begin{align}
\Theta=\psi^\mu p_\mu + F_{\mu_1\cdots \mu_7}\psi^{\mu_1}\cdots \psi^{\mu_7}+ \zeta F_{\mu_1\cdots \mu_4}\psi^{\mu_1}\cdots \psi^{\mu_4}\,.
\end{align}
Therefore $\Theta$ depends on a 4-form and a 7-form which we will respectively write $F_4$ and $F_7$. Remarkably, the condition $(\Theta,\Theta)=0$ implies the Bianchi identity and field equation for the $C$-field of 11d SUGRA (or, rather, their components along $M$):
\begin{align}
(\Theta,\Theta)=0\iff d F_4=0\,,\quad  d F_7 + \frac{1}{2}F_4 F_4=0\,.
\end{align}

The upshot is that the exact Courant, M2 and M5 algebroids are classified by classes which directly correspond to the coupling of the field strengths of the $b$-field, $C$-field and its dual respectively to 1-branes, 2-branes and 5-branes (for the last one observe that $F_7$ is naturally integrated against a 7-fold which leads to an electric $\tilde C$ Wess-Zumino coupling in the presence of a 6-fold boundary). In fact this can be made rather direct using the AKSZ construction \cite{Alexandrov:1995kv}, which associates a topological field theory of a $(p-1)$-brane to any degree $p$ NPQ manifold. This is explained in \cite{Arvanitakis:2018cyo} where the AKSZ construction is found to directly connect the M5 brane construction here to 7-dimensional Chern-Simons theory and the gauge theory on the M5 brane.

\section*{Acknowledgements}
I am supported by the EPSRC programme grant ``New Geometric Structures from String Theory'' (EP/K034456/1)

\bibliographystyle{JHEP}
\bibliography{NewBib}
\end{document}